\documentclass[letter,twocolumn]{jpsj3}
\usepackage{graphicx}
\usepackage{txfonts}
\usepackage{color}
\usepackage{amssymb}
\usepackage{ulem}

\title{Quantized $\Delta S=2$ Excitation Spectra by Confinement in an $S=1$ Spin Chain}

\author{Takafumi Suzuki and Sei-ichiro Suga}
\inst{Graduate School of Engineering, University of Hyogo, Himeji, Hyogo 671-2280, Japan} 

\abst{
We calculate the dynamical spin-structure factor of the $S=1$ Ising spin chain with negative single-ion anisotropy in magnetic fields using the infinite time-evolving-block-decimation algorithm. 
We show that when a transverse magnetic field is applied, both the $\Delta S=2$ excitation continuum and one-magnon mode appear in the low-lying excitation. 
When a longitudinal magnetic field is further applied, the excitation continuum changes into quantized excitation spectra. 
The quantized $\Delta S=2$ excitation spectra originate from the confinement of two domain walls, each of which carries $\Delta S=1$. The quantized excitation energies are explained by the negative zeros in the Airy function. }

\begin{document}
\maketitle
In a seminal study for the $S=1/2$ ferromagnetic (FM) Ising spin chain in weak transverse and longitudinal magnetic fields, it was unveiled that quantized excitation spectra appear by the confinement of the domain-wall excitation \cite{McCoy}. 
The lowest excitation of the $S=1/2$ FM Ising spin chain is achieved by flipping the spins of arbitrary lengths. Each domain wall of both ends carries $\Delta S=1/2$, namely spinon. 
In the transverse magnetic field, two spinons travel in the chain and compose an excitation continuum. When the longitudinal magnetic field is further weakly applied, it works as a confinement potential between two spinons, yielding the quantized excitation spectra whose excitation energies are described by the series of negative zeros in the Airy function (NZAF)~\cite{McCoy}.
The quantized spectra described by the NZAF have been confirmed in the inelastic neutron scattering (INS) experiment \cite{RColdea} on ${\rm CoNb_2O_6}$, which is a FM Ising-spin-chain compound.

The quantized excitation spectra of the quasi-one-dimensional (q1D) $S=1/2$ antiferromagnetic (AF) Heisenberg spin system have also been elucidated \cite{HShiba}. 
Recently, the quantized excitation spectra of the q1D $S=1/2$ AF Ising-like XXZ magnets, (Ba/Sr)Co$_2$V$_2$O$_8$ \cite{SKimura}, have been observed in the INS experiments \cite{BGrenier,QFaure}. 
In these q1D AF spin systems, the excitation continuum originating from spinons appears in the low-lying excitation above the Ne\'{e}l temperature ($T_N$). Below $T_N$, an effective staggered field  that works as a confinement potential is induced in the spin chain.  
Thus, the observed quantized excitation energies \cite{BGrenier,QFaure} are explained by the NZAF in the similar manner to the discussion of the FM Ising-spin chain. 
This scenario has been further applied to systems in which the excitation continuum is generated by quasiparticles. The $S=1$ AF Heisenberg chain is a typical system where the excitation continuum is generated by multimagnons \cite{SRWhite2,IAffleck,SYamamoto,MDPHorton,MTakahashi,FHLEssler}. 
We calculated the dynamical spin structure factor (DSF) of the q1D $S=1$ AF Heisenberg system with single-ion anisotropy, and demonstrated that quantized excitation spectra appear \cite{Suzuki2018}. The quantized excitation energies are well described by the NZAF, when the single-ion anisotropy is negatively strong.

In INS experiments, neutrons are scattered by changing spins in the target systems by $\Delta S=1$, which makes it possible to detect the low-lying excitation that composes the excitation continuum.  
The quantized excitation spectra in $S=1/2$ spin systems have been observed in the INS experiments \cite{RColdea,BGrenier,QFaure}, because the $\Delta S=1$ excitation generates the original excitation continuum. 
The quantized excitation spectra in the q1D $S=1$ AF Heisenberg system with single-ion anisotropy are possibly observed in the INS experiment, because its excitation continuum is generated by the multimagnons with each magnon carrying $\Delta S=1$ \cite{Suzuki2018}. 
In this Letter, we show that the quantized $\Delta S=2$ excitation spectra are generated in the DSF of an $S=1$ spin chain. 
The quantized $\Delta S=2$ excitation spectra provide observations in the INS experiment. 

We consider the $S=1$ FM Ising spin chain with single-ion anisotropy in weak transverse and longitudinal magnetic fields. The Hamiltonian is written as
\begin{eqnarray}
{\mathcal H} = J  {\displaystyle \sum_{i}} S_i^z S_{i+1}^z  + D_z {\displaystyle \sum_{i}} \left(S_{i}^z \right)^2 
- H_x {\displaystyle \sum_{i}} {S_i^x} - H_z {\displaystyle \sum_{i}} {S_i^z},    
\end{eqnarray}
where $J<0$ and the single-ion anisotropy $D_z<0$. 
In the following calculations, we focus on the system whose ground state is in the FM state for  $H_z=0$.

We apply the infinite time-evolving-block-decimation algorithm\cite{GVidal,ROrus} to calculate the DSF. 
The DFS is defined as $S^{\mu\mu}(q_x,\omega)=\pi^{-1}{\rm Im} \int \int i\langle {S_x}^{\mu}(t){S_0}^{\mu}(0)\rangle e^{-iq_xx-i(\omega-e_g)t}dxdt$, 
where $\mu=x,y,z$ and $e_g$ is the ground-state energy. 
The details of the numerical techniques have been discussed in Ref. \cite{HNPhien}.
To reduce numerical noise, we combine the Gaussian filtering method~\cite{SRWhite} with the Fourier transformation.
In the following calculations, we set  $\chi_{\rm max}=80$ and $N=200$, where $\chi_{\rm max}$ is the maximum bond dimension for tensors comprising the wave function and $N$ is the real-space window size for the Fourier transformation, respectively.

\begin{figure}[htb]
\begin{center}
\includegraphics[bb=0 0 1250 1600, scale=0.18]{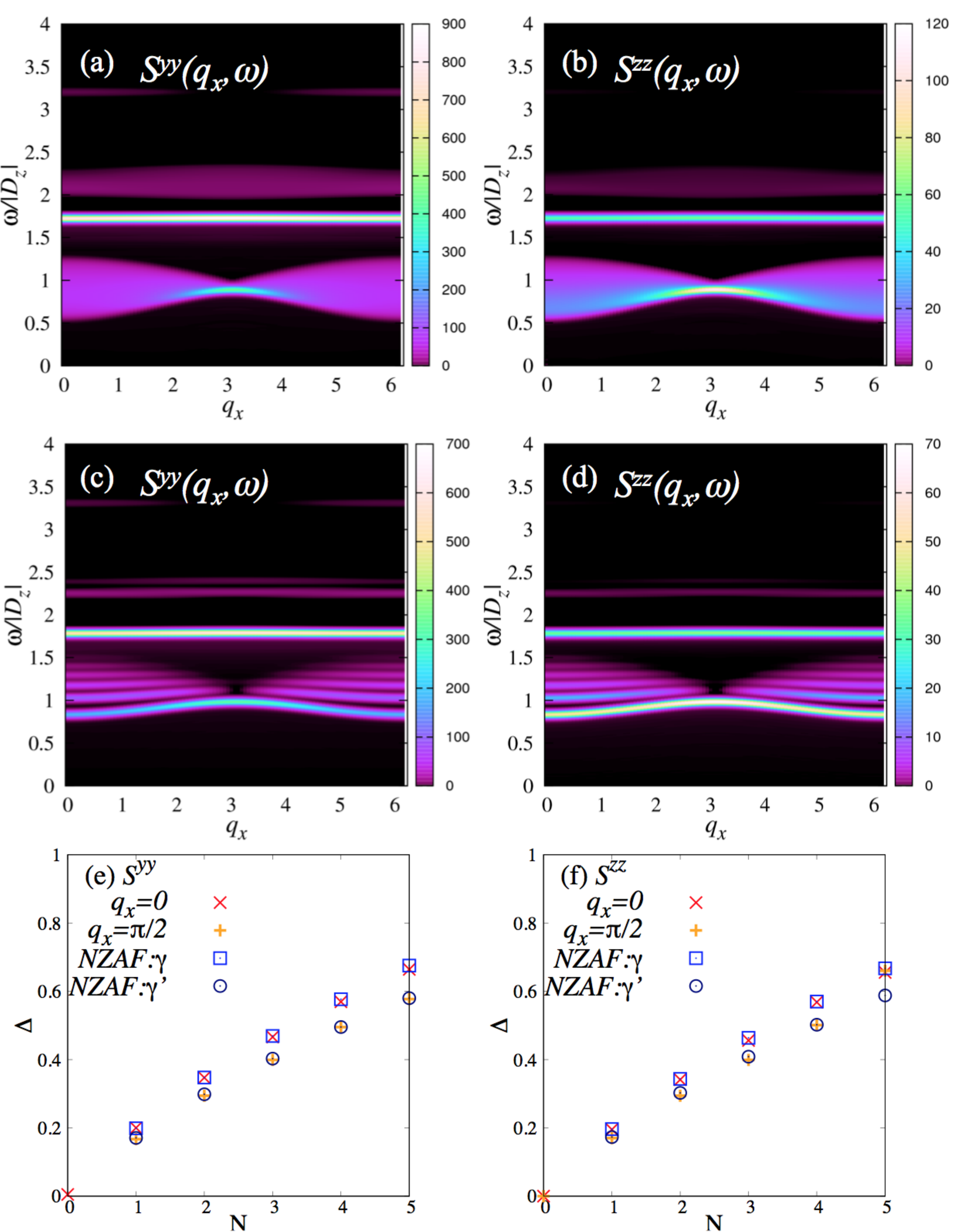}
\hspace{0pc}
\vspace{-0.5pc}
\caption{\label{fig2} (Color online) For $(J,D_z,H_x)=(-0.25,-1,0.45)$; (a) and (c) $S^{yy}(q_x,\omega)$ for $H_z=0$ and $0.05$, respectively, and (b) and (d) $S^{zz}(q_x,\omega)$ for $H_z=0$ and $0.05$, respectively.
The quantized excitation energies, $\Delta$, at $q_x=0$ and $\pi/2$  of (e) $S^{yy}(q_x,\omega)$ and (f) $S^{zz}(q_x,\omega)$ are compared with the NZAF. $\Delta$ is measured from the lowest quantized energy. 
The horizontal axis denotes the level of the quantized excitation energies. 
The NZAF are scaled using a constant factor $\gamma$ to fit the first quantized excitation energy $\Delta(N=1)$. Thus, $\Delta=\gamma z_i$, where $z_i$ indicates the negative zeros of the Airy function. 
The constant $\gamma$ is $\approx 0.085$ in both $S^{yy}(q_x=0,\omega)$ and $S^{zz}(q_x=0,\omega)$. 
In $S^{yy}(q_x=\pi/2,\omega)$ and $S^{zz}(q_x=\pi/2,\omega)$, $\gamma \approx 0.073$ is adopted. 
}
\end{center}
\end{figure}

In Figs. \ref{fig2}(a) and \ref{fig2}(c), $S^{yy}(q,\omega)$ is shown for $(J, D_z, H_x)=(-0.25, -1, 0.45)$. 
Note that the same behavior is observed for $S^{xx}(q,\omega)$.  
We investigate the phase transition between the FM and paramagnetic states driven by $H_x$. For $(J, D_z)=(-0.25, -1)$ and $H_z=0$, we confirm that the phase transition occurs at $H_x \approx 0.75$. 
For $H_z=0$, the excitation continuum appears below $\omega/|D_z|<1.5$. When $H_z$ is switched on, the excitation continuum changes to the quantized spectra. 
We discuss the feature of the quantized spectra in $S^{yy}(q_x,\omega)$.
For small $H_x$ and $H_z$, we divide ${\mathcal H}={\mathcal H}_0+{\mathcal H}_1$, where ${\mathcal H}_0 = J  {\sum_{i}} S_i^z S_{i+1}^z  + D_z {\sum_{i}} \left(S_{i}^z \right)^2$ and ${\mathcal H}_1=- H_x {\sum_{i}} {S_i^x} - H_z {\sum_{i}} {S_i^z}$. 
The ground state of ${\mathcal H}_0$ is the fully polarized state expressed by $\psi^{+}_{\rm GS} = |\cdots $+++++++$ \cdots \rangle$ or $\psi^{-}_{\rm GS}|\cdots -------\cdots \rangle$, where $+$, $0$, and $-$ in the ket denote $S^z=1$, $0$, and $-1$, respectively. 
In the following discussion, we adopt the former ground state, $\psi^{+}_{\rm GS}$. 
The low-lying excitation in $S^{yy}(q_x,\omega)$ is described by the dynamics of the excited  state whose initial state is $S_i^{y}\psi^{+}_{\rm GS} \propto S_i^{-}\psi^{+}_{\rm GS}$. Thus, this initial state is interpreted as a one-magnon state, $|\cdots $+++0+++$ \cdots \rangle$. 
As the energy cost to create $S_i^{-}\psi^{+}_{\rm GS}$ is approximately $\omega = |D_z|+2|J|$,  the isolated mode by $S_i^{-}\psi^{+}_{\rm GS}$ appears at $\omega \approx |D_z|+2|J|$ for ${\mathcal H}_1=0$. 
In Figs. \ref{fig2}(a) and \ref{fig2}(c), an almost dispersionless mode with a large intensity appears at $\omega/|D_z| \approx 1+2|J/D_z|=1.5$. 
Therefore, this mode is considered to be the one-magnon mode with $\Delta S=1$.

\begin{figure}[thb]
\begin{center}
\includegraphics[bb=0 0 1250 1600, scale=0.18]{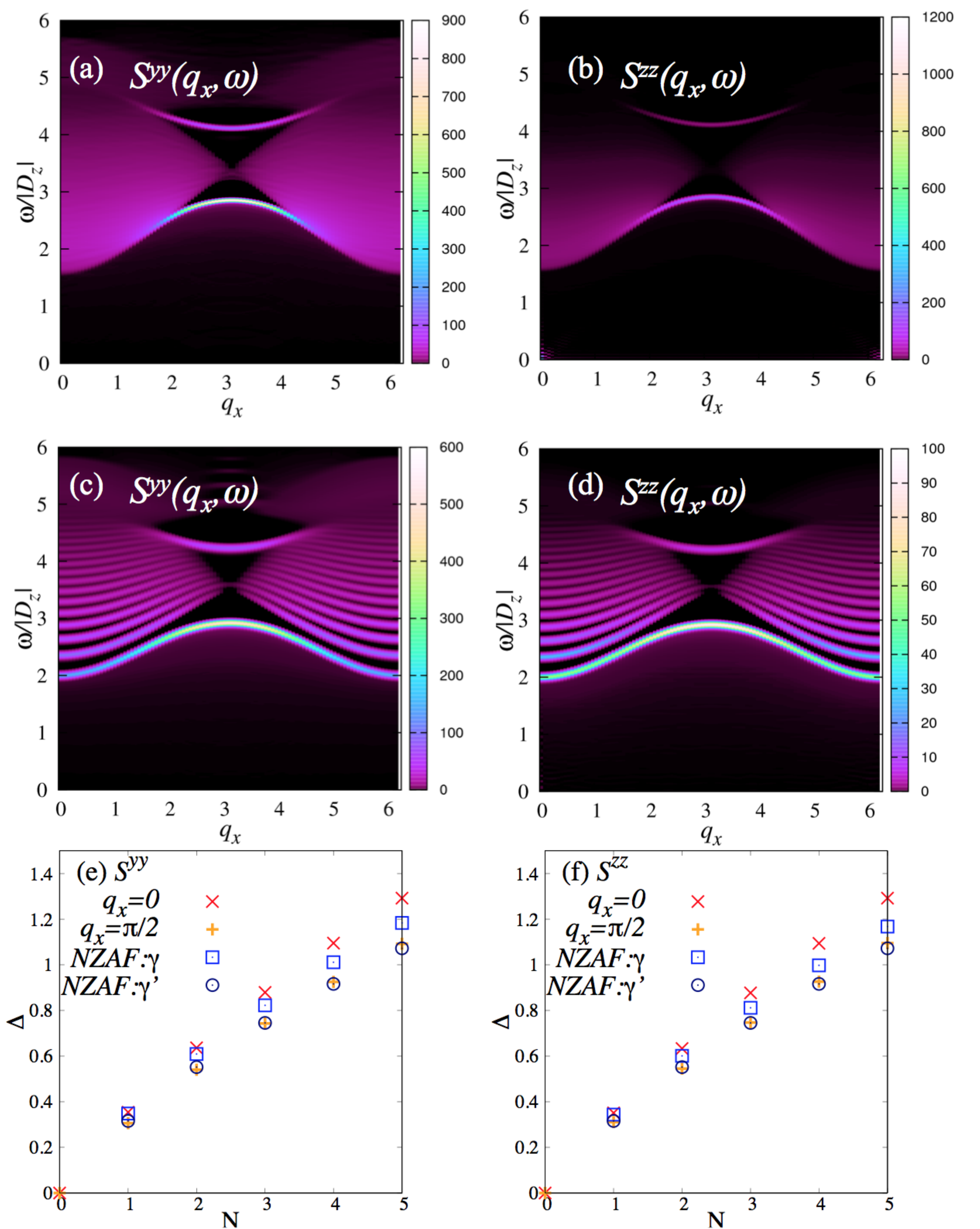}
\hspace{0pc}
\vspace{-0.5pc}
\caption{\label{fig4} (Color online) For $(J,D_z,H_x)=(-1, -1, 1.25)$; (a) and (c) $S^{yy}(q_x,\omega)$ at $H_z=0$ and $0.05$, respectively, and (b) and (d)
$S^{zz}(q_x,\omega)$ at $H_z=0$ and $0.05$, respectively. The quantized excitation energies, $\Delta$, of (e) $S^{yy}(q_x,\omega)$ and (f) $S^{zz}(q_x,\omega)$, at $q_x=0$ and $\pi/2$  are compared with the NZAF. $\Delta$ is measured from the lowest quantized energy. 
The horizontal axis denotes the level of the quantized excitation energies. 
The NZAF are scaled using a constant factor, $\gamma$, to fit the first quantized excitation energy $\Delta(N=1)$. Thus, $\Delta=\gamma z_i$, where $z_i$ indicates the negative zeros of the Airy function. 
$\gamma \approx 0.145$ in both $S^{yy}(q_x=0,\omega)$ and $S^{zz}(q_x=0,\omega)$.
For $S^{yy}(q_x=\pi/2,\omega)$ and $S^{zz}(q_x=\pi/2,\omega)$, $\gamma \approx 0.135$ is adopted. 
}
\end{center}
\end{figure}

When the transverse magnetic field, $- H_x {\sum_{i}} {S_i^x}$, in ${\mathcal H}_1$ acts on the site with $S^z=0$, a further excited state with $S^z=-1$ appears because $S_i^{x}S_i^{y}\psi^{+}_{\rm GS} \propto (S_i^{-})^2\psi^{+}_{\rm GS}$, yielding $| \cdots$+++$-$+++$\cdots \rangle$. 
This excited state carries $\Delta S=2$.   
The energy cost for creating the $\Delta S=2$ excited state is $4|J|$, which is the same as that for creating the excited state with an $S^z=-1$ domain, $|\cdots$ +++$-\cdots-$+++$ \cdots \rangle$. Therefore, the domain comprising the $S^z=-1$ sites develops by the second-order process with respect to $- H_x {\sum_{i}} {S_i^x}$, yielding an excitation continuum, as shown in Fig. \ref{fig2}(a). 
Using the analogy of the domain-wall excitation in the $S=1/2$ FM Ising spin chain, we call this elementary excitation ``$\Delta S=1$ spinon'' because each domain wall carries $\Delta S=1$. 
Both the one-magnon mode and $\Delta S=2$ excitation continuum appear, as shown in Fig. \ref{fig2}(a). This is a characteristic feature of the present $S=1$ systems. 
In contrast, only the spinon continuum appears in the low-lying excitation of the $S=1/2$ FM Ising spin chain in the transverse magnetic field \cite{McCoy}.   
When $2|J|<|D_z|$, the excitation energy of $(S_i^{-})^2\psi^{+}_{\rm GS}$ is less than that of $S_i^{-}\psi^{+}_{\rm GS}$. 
The present parameters, $(J, D_z)=(-0.25, -1)$, satisfy this condition. Thus, an excitation continuum with the band center at $\omega/|D_z| \approx 4|J/D_z|=1.0$ interestingly appears below the one-magnon mode. 
On the other hand, in Figs. \ref{fig4}(a)--\ref{fig4}(d), $S^{yy}(q_x,\omega)$ and $S^{zz}(q_x,\omega)$ are shown for $(J, D_z, H_x)=(-1, -1, 1.25)$, where $2|J|>|D_z|$, contrary to the aforementioned discussion. 
In this case, the excitation continuum and quantized excitation spectra appear above the one-magnon mode.

When the longitudinal magnetic field, $- H_z {\sum_{i}} {S_i^z}$, in ${\mathcal H}_1$ is switched on to the $\Delta S=2$ excitation continuum, the excitation continuum changes to the quantized spectra, as shown in Fig. \ref{fig2}(c). 
In Fig. \ref{fig2}(e), we compare the quantized excitation energies, $\Delta$, with the NZAF, where $\Delta$ is measured from the lowest quantized energies at $q_x=0$ and $\pi/2$.  
The quantized excitation energies agree well with the NZAF. 
For other $q_x$'s, except for $q_x \approx \pi$ where the excitation continuum shrinks, the quantized excitation energies also agree with the NZAF. Thus, the longitudinal magnetic field is confirmed to act as a confinement potential for two $\Delta S=1$ spinons.

$S^{zz}(q_x,\omega)$ is shown in Figs. \ref{fig2}(b) and \ref{fig2}(d). The initial state of the excited state in $S^{zz}(q_x,\omega)$ is $S_i^{z}\psi^{+}_{\rm GS}$. 
When the transverse magnetic field, $- H_x {\sum_{i}} {S_i^x}$, in ${\mathcal H}_1$ is applied, it changes the $S^z=1$ state into the $S^z=0$ state in the first-order process and it changes the $S^z=0$ state into the $S^z=-1$ state in the second-order process. 
By repeating the second-order process, the domain comprising $S^z=-1$ sites develops and the excitation continuum appears, as shown in  Fig. \ref{fig2}(b). In other words, two $\Delta S=1$ spinons are also realized in  $S^{zz}(q_x,\omega)$. 
When the longitudinal magnetic field, $- H_z {\sum_{i}} {S_i^z}$, is further applied, the excitation continuum changes into the quantized spectra and their quantized energies agree well with the NZAF as shown in Figs. \ref{fig2}(d) and \ref{fig2}(f).
Therefore, the longitudinal field, $- H_z {\sum_{i}} {S_i^z}$, also works as a confinement potential for two $\Delta S=1$ spinons in $S^{zz}(q_x,\omega)$.

In Figs. \ref{fig4}(a)--\ref{fig4}(d), $S^{yy}(q_x,\omega)$ and $S^{zz}(q_x,\omega)$ for $(J, D_z, H_x)=(-1, -1, 1.25)$ are shown. 
We confirm that the phase transition between the FM and paramagnetic states occurs at $H_x \approx 1.9$ for $(J, D_z)=(-1, -1)$ and $H_z=0$. 
The excitation continuum in $H_z=0$ and quantized spectra in $H_z \neq 0$ appear above the one-magnon mode because of the condition, $2|J|>|D_z|$, as discussed before. 
The quantized excitation energies are compared with the NZAF in Figs. \ref{fig4}(e) and \ref{fig4}(f). We find that they are explained by the NZAF, which means that the confinement of two $\Delta S=1$ spinons occurs because of the longitudinal magnetic field in $S^{yy}(q_x,\omega)$ and $S^{zz}(q_x,\omega)$.   
The quantized excitation energies at $q_x=0$ deviate slightly from the NZAF in the higher energies. This deviation is probably caused by the large value of $H_x$.

In summary, we have demonstrated that the quantized $\Delta S=2$ excitation spectra, caused by the two $\Delta S=1$ spinon confinement, occur in addition to the one-magnon mode in the $S=1$ FM Ising chain with negative single-ion anisotropy under transverse and longitudinal magnetic fields.  
In the present system, the $\Delta S=2$ excitation continuum appears because the transverse magnetic field, $- H_x {\sum_{i}} {S_i^x}$, makes two $\Delta S=1$ spinons propagate in the DSF.  
On the other hand, in the $S=1$ AF Heisenberg chain with negative single-ion anisotropy, the excitation continuum originates from $\Delta S=1$ magnons that are propagated by the transverse components ${\sum_{i}} \left( S_i^xS_{i+1}^x + S_i^yS_{i+1}^y \right)$ \cite{Suzuki2018}. 
In general, it is difficult to observe $\Delta S=2$ excitations by INS experiments. 
However, in this study we have shown that the quantized $\Delta S=2$ excitation spectra appear in the DSF, which is evidence of two $\Delta S=1$ spinon confinement. 
Since the DSF can be observed by INS experiments, we expect the present quantized $\Delta S=2$ excitation spectra to be detected by INS experiments.

\begin{acknowledgment}
This work was supported by the CDMSI, CBSM2, and KAKENHI (Grant No. 16K17751) from MEXT, Japan. 
T.S. thanks the computational resources of the K computer provided by the RIKEN AICS through the HPCI System Research Project (hp170262, hp170263, hp180170, and hp180225). 
We are also grateful for the numerical resources at the ISSP Supercomputer Center at the University of Tokyo  
and the Research Center for Nano-Micro Structure Science and Engineering at University of Hyogo.
\end{acknowledgment}

\end{document}